\documentclass[twocolumn,aps,prc,showpacs,floatfix]{revtex4}
\usepackage{graphicx}
\usepackage{dcolumn}
\usepackage{bm}
\usepackage{CJK}

\begin{document}
\begin{CJK*}{GBK}{song}

\preprint{}
\title{Effects of symmetry energy in the reaction $^{40}$Ca+$^{124}$Sn@140 MeV/nucleon}
\author{ Zhang Fang, Hu Bi-Tao}
\affiliation{School of Nuclear Science and Technology, Lanzhou
University, Lanzhou 730000, China}
\author{ Yong Gao-Chan and Zuo Wei }\affiliation{Institute of
Modern Physics, Chinese Academy of Sciences, Lanzhou 730000,
China}


\begin{abstract}
The density dependent symmetry energy is a hot topic in nuclear
physics. Many laboratories in the word are planning to do related
experiments to probe the symmetry energy. Based on the
semiclassical Boltzmann-Uehling-Uhlenbeck (BUU) transport model,
we studied the effects of nuclear symmetry energy in the central
reaction $^{40}$Ca+$^{124}$Sn at 140 MeV/nucleon in the laboratory
system. It is found that the rapidity distribution of free
nucleon's neutron to proton ratio is sensitive to the symmetry
energy, especially at large rapidities. The free \emph{n/p} ratios
at small or large rapidities may reflect high or low density
behavior of nuclear symmetry energy. To probe the density
dependence of nuclear symmetry energy, it is better to give the
kinetic distribution and the rapidity distribution of emitted
nucleons at the same time.
\end{abstract}

 \maketitle

\section{Introduction}
After about three decades of intensive efforts in both nuclear
experiments and theories, the equation of state (EOS) of isospin
symmetric nuclear matter is now relatively well determined mainly
by studying collective flows in heavy-ion collisions and nuclear
giant monopole resonances \cite{pd02,youngblood99}. The major
remaining uncertainty about the EOS of symmetric nuclear matter is
due to our poor knowledge about the density dependence of the
nuclear symmetry energy
\cite{xiao09,LCK08,Bar05,pd02,pie04,colo04}. Therefore, the new
challenge is to determine the EOS of isospin asymmetric nuclear
matter, especially the density dependence of the nuclear symmetry
energy. Besides the great interests in nuclear physics, the EOS of
asymmetric nuclear matter is also crucial in many astrophysical
processes, especially in connection with the structure of neutron
stars and the dynamical evolution of proton-neutron stars
\cite{mk94}. Considerable progress has been made recently in
determining the density dependence of the nuclear symmetry energy
around the normal nuclear matter density. However, much more work
is still needed to probe the high-density behavior of the nuclear
symmetry energy. Currently, to pin down the symmetry energy, the
National Superconducting Cyclotron Laboratory (NSCL) at Michigan
State University, the Gesellschaft fuer Schwerionenforschung (GSI)
at Darmstadt, the Rikagaku Kenkyusho (RIKEN, The Institute of
Physical and Chemical Research) of Japan, and the Cooler Storage
Ring (CSR) in Lanzhou, are planning to do related experiments.
Effects of the symmetry energy on the pre-equilibrium
neutron/proton ratio in heavy ion collisions have been analyzed
with many authors. For example, calculations by Zhang et
al.\cite{zhang} within the Improved Quantum Molecular
Dynamics(ImQMD) model using two different density-dependent
symmetry-energy functions, the results of neutron to proton ratios
are sensitive to the density dependence of the symmetry energy.
Within the transport model Lanzhou Quantum Molecular
Dynamics(LQMD), Feng \cite{feng} investigated the single and
double neutron to proton ratio of free nucleons with different
collision centralities and effective mass splitting. Recently, Ma
et al.\cite{ma} studied the high-density behavior of the symmetry
energy by using single and double ratios of neutrons to protons
within Isospin Quantum Molecular Dynamics(IQMD) model, they
confirmed that it is possible to study the high-density behavior
of symmetry energy by using the neutron-to-proton ratio from free
nucleons. In the present work we give our results of the effects
of symmetry energy on free neutron to proton ratio as a function
of rapidity in the $^{40}$Ca+$^{124}$Sn at 140 MeV/nucleon. We
find that nucleon emissions at low-angle and large rapidities
regions are suitable to be used to probe the effect of symmetry
energy. The free \emph{n/p} ratios at large or small rapidities
may reflect low or high density behavior of symmetry energy.

\section{THE IBUU04 TRANSPORT MODEL}
Our present study is based on the transport model IBUU04. In this
semi-classical model  besides nucleons, \emph{$\Delta $} and
$N^{\ast }$ resonances as well as pions and their
isospin-dependent dynamics are included, the experimental
free-space nucleon-nucleon (NN) scattering cross sections and the
in-medium NN cross sections can be used optionally. By using the
relativistic mean field theory, the initial neutron and proton
density distributions of the projectile and target are obtained.
The isospin dependent phase-space distribution functions of the
particles involved are solved by using the test-particle method
numerically. The isospin-dependence of Pauli blockings for
fermions is also considered. More details can be found in Refs.
 \cite{li05,li03,li04,das03,l04,yong06}. In the present work, we use the
isospin-dependent in-medium NN elastic cross sections from the
scaling model according to nucleon effective masses \cite{li05}.
In the IBUU04 transport model, the most important input is the
momentum- and isospin-dependent single nucleon potential, we use a
single nucleon potential derived within the Hartree-Fock approach
using a modified Gogny effective interaction \cite{das03}, the
momentum-dependent single nucleon potential (MDI) adopted here
is:%
\begin{eqnarray}
U(\rho ,\delta ,\mathbf{p},\tau ) &=&A_{u}(x)\frac{\rho _{\tau ^{\prime }}}{%
\rho _{0}}+A_{l}(x)\frac{\rho _{\tau }}{\rho _{0}}  \nonumber \\
&&+B(\frac{\rho }{\rho _{0}})^{\sigma }(1-x\delta ^{2})-8x\tau \frac{B}{%
\sigma +1}\frac{\rho ^{\sigma -1}}{\rho _{0}^{\sigma }}\delta \rho _{\tau
^{\prime }}  \nonumber \\
&&+\frac{2C_{\tau ,\tau }}{\rho _{0}}\int d^{3}\mathbf{p}^{\prime }\frac{%
f_{\tau }(\mathbf{r},\mathbf{p}^{\prime })}{1+(\mathbf{p}-\mathbf{p}^{\prime
})^{2}/\Lambda ^{2}}  \nonumber \\
&&+\frac{2C_{\tau ,\tau ^{\prime }}}{\rho _{0}}\int d^{3}\mathbf{p}^{\prime }%
\frac{f_{\tau ^{\prime }}(\mathbf{r},\mathbf{p}^{\prime })}{1+(\mathbf{p}-%
\mathbf{p}^{\prime })^{2}/\Lambda ^{2}}.  \label{potential}
\end{eqnarray}%
In the above equation, $\delta =(\rho _{n}-\rho _{p})/(\rho
_{n}+\rho _{p})$ is the isospin asymmetry parameter, $\rho =\rho
_{n}+\rho _{p}$ is the baryon density and $\rho _{n},\rho _{p}$
are the neutron and proton densities, respectively. $\tau
=1/2(-1/2)$ for neutron (proton) and $\tau \neq \tau ^{\prime }$,
$\sigma =4/3$, $f_{\tau }(\mathbf{r},\mathbf{p})$ is the
phase-space distribution function at coordinate $\mathbf{r}$ and
momentum $\mathbf{p}$. The parameters $A_{u}(x),A_{l}(x),B,C_{\tau
,\tau }$, $C_{\tau ,\tau ^{\prime }}$ and $\Lambda $ were set by
reproducing the momentum-dependent potential $U(\rho ,\delta
,\mathbf{p},\tau )$ predicted by the Gogny Hartree-Fock and/or the
Brueckner-Hartree-Fock calculations. The momentum-dependence of
the symmetry potential stems from the different interaction
strength parameters $C_{\tau,\tau'}$ and $C_{\tau,\tau}$ for a
nucleon of isospin $\tau$ interacting, respectively, with unlike
and like nucleons in the background fields, more specifically,
$C_{unlike}=-103.4$ MeV while $C_{like}=-11.7$ MeV. The parameters
$A_{u}(x)$and $A_{l}(x)$ depend on the $x$ parameter according to
$Au(x)=-95.98-x\frac{2B}{\sigma+1}$ and $A_{l}(x) =
-120.57+x\frac{2B}{\sigma+1}$. The saturation properties of
symmetric nuclear matter and the symmetry
energy of about $32$ MeV at normal nuclear matter density $\rho _{0}=0.16$ fm%
$^{-3}$. The incompressibility of symmetric nuclear matter at
normal density is set to be $211$ MeV. According to essentially
all microscopic model calculations, the EOS for isospin asymmetric
nuclear matter can be expressed as
\begin{equation}
E(\rho ,\delta )=E(\rho ,0)+E_{\text{sym}}(\rho )\delta ^{2}+\mathcal{O}%
(\delta ^{4}),
\end{equation}%
where $E(\rho ,0)$ and $E_{\text{sym}}(\rho )$ are the energy per
nucleon of symmetric nuclear matter and nuclear symmetry energy,
respectively. For a given value $x$, with the single particle
potential $U(\rho ,\delta,\mathbf{p},\tau )$, one can readily calculate the symmetry energy $E_{\text{sym}%
}(\rho )$ as a function of density.

\section{Results and discussions}

\begin{figure}[th]
\begin{center}
\includegraphics[width=0.5\textwidth]{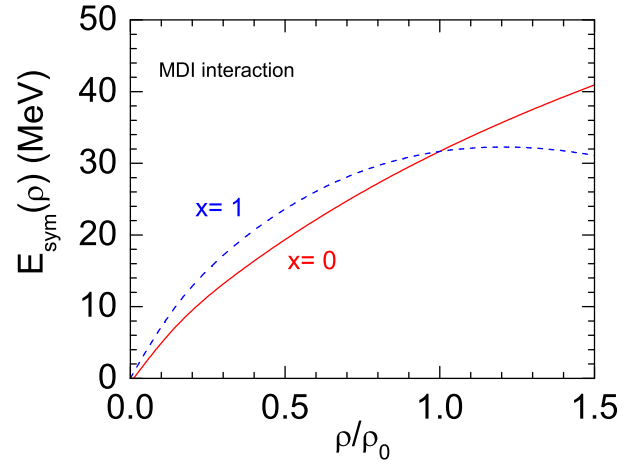}
\end{center}
\caption{{\protect\small (Color online) Density dependence of
nuclear symmetry energy with parameters $x= 1, 0$, respectively.}}
\label{Esym}
\end{figure}
\begin{figure}[th]
\begin{center}
\includegraphics[width=0.5\textwidth]{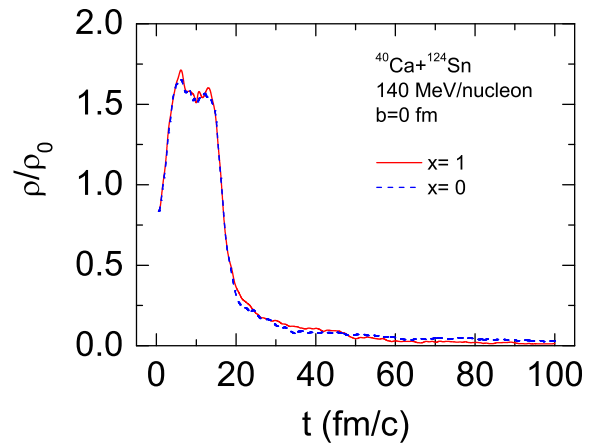}
\end{center}
\caption{{\protect\small (Color online) Maximal baryon density
reached in the central reaction $^{40}$Ca+$^{124}$Sn at 140
MeV/nucleon.}} \label{den}
\end{figure}
Fig.\ \ref{Esym} shows the density dependence of nuclear symmetry
energy with parameter $x=1, 0$, respectively. As discussed in the
previous part, the single particle potential used has an $x$
parameter, different specific $x$ parameter denotes different
density dependent symmetry energy. For the central reaction
$^{40}$Ca+$^{124}$Sn at 140 MeV/nucleon, the maximal density
reached is about 1.5 times saturation density as shown in Fig.\
\ref{den}. We therefore only show the low density's symmetry
energy as a function of density. From Fig.\ \ref{Esym}, we can
also see that the low density behaviors of nuclear symmetry energy
separate from each other with different $x$ parameters. At the
saturation point there is a cross and then they separate from each
other again. At lower densities, the value of symmetry energy of
$x=0$ is lower than that of $x=1$. But at high densities, the
value of symmetry energy of $x=0$ is higher than that of $x=1$.
From Fig.\ \ref{den}, we can see that at the reaction time $t=50$
fm/c, the reaction almost ends.

\begin{figure}[th]
\begin{center}
\includegraphics[width=0.5\textwidth]{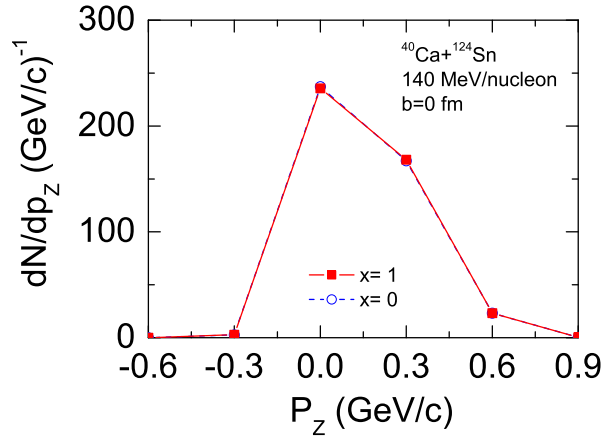}
\end{center}
\caption{{\protect\small (Color online) Longitudinal momentum
($p_{z}$) distribution of emitted free nucleons in laboratory
system in the central reaction $^{40}$Ca+$^{124}$Sn at 140
MeV/nucleon.}} \label{pd}
\end{figure}
\begin{figure}[th]
\begin{center}
\includegraphics[width=0.5\textwidth]{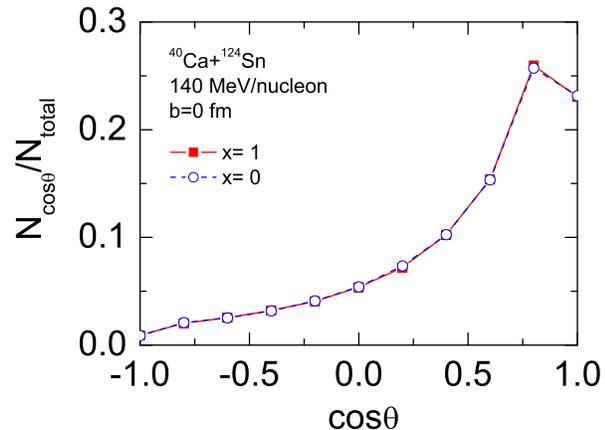}
\end{center}
\caption{{\protect\small (Color online) Angle distribution of
relative emitted number ($N\cos\theta/N_{total}$) of the free
nucleons in laboratory system in the central reaction
$^{40}$Ca+$^{124}$Sn at 140 MeV/nucleon.}} \label{cosd}
\end{figure}
Fig.\ \ref{pd} shows the longitudinal momentum ($p_{z}$)
distribution of emitted free nucleons in laboratory system in the
central reaction $^{40}$Ca+$^{124}$Sn at 140 MeV/nucleon. In our
work, free nucleons are identified as those having local baryon
densities less than $\rho=\frac{1}{8}\rho_{0}$. We can first see
that more nucleons are emitted at positive momentum part than at
the negative part. Also one can see that longitudinal momentum
($p_{z}$) distribution of emitted free nucleons is insensitive to
the symmetry energy since the effects of symmetry energy are
always very small. The maximal longitudinal momentum at positive
momentum part reached is about 0.9 GeV/c, but the maximal
longitudinal momentum at negative momentum part reached is only
about 0.3 GeV/c. From this plot we can clearly see that most
emissions are at very small longitudinal momentum ($p_{z}$),
denoting that more nucleons may emitted perpendicularly to the
reaction plane. Experimentally, one always needs to know the angle
distribution of the emitted number of probed nucleons in the
reaction. This is because one usually can not probe the emitted
particles in $4\pi$ directions. For this purpose, we plot Fig.\
\ref{cosd}, the angle distribution of relative emitted number
($N\cos\theta/N_{total}$) of the free nucleons in laboratory
system in the central reaction $^{40}$Ca+$^{124}$Sn at 140
MeV/nucleon. From this plot, we can clearly see that whether for
$x=1$ or $x=0$ nucleons are inclined to low-angle emission. It is
thus suitable to probe the emitted nucleons in the low-angle
regions.

\begin{figure}[th]
\begin{center}
\includegraphics[width=0.5\textwidth]{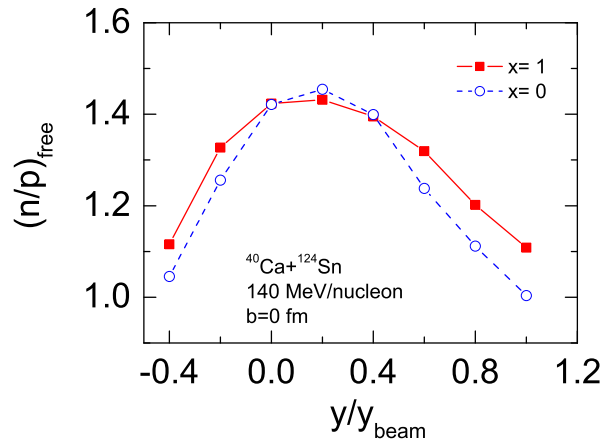}
\end{center}
\caption{{\protect\small (Color online) Neutron to proton ratio
(n/p) of the emitted free nucleons as a function of reduced
rapidity with $x=1$ and 0 in laboratory system in the central
reaction $^{40}$Ca+$^{124}$Sn at 140 MeV/nucleon.}} \label{rnp}
\end{figure}
\begin{figure}[th]
\begin{center}
\includegraphics[width=0.5\textwidth]{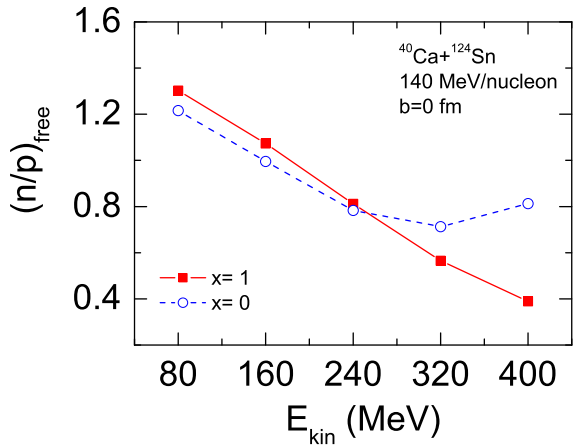}
\end{center}
\caption{{\protect\small (Color online) Neutron to proton ratio
(n/p) of the emitted free nucleons as a function of kinetic energy
with x= 1 and 0 in laboratory system in the central reaction
$^{40}$Ca+$^{124}$Sn at 140 MeV/nucleon.}} \label{rnp2}
\end{figure}
To see the effects of symmetry energy on emitted free neutron to
proton ratio, we provide Fig.\ \ref{rnp}, the neutron to proton
ratio (\emph{n/p}) of the emitted free nucleons as a function of
reduced rapidity with $x=1$ and 0 in laboratory system in the
central reaction $^{40}$Ca+$^{124}$Sn at 140 MeV/nucleon. From
this plot we can see that at large rapidities, effects of symmetry
energy are clearly shown on the neutron to proton ratio. The soft
symmetry energy ($x=1$) corresponds large \emph{n/p} ratio than
the stiff case ($x=0$), the \emph{n/p} ratio thus reflects the
low-density behavior of nuclear symmetry energy. Around low
rapidity $y/y_{beam}$= 0.2, the stiff symmetry energy corresponds
large \emph{n/p} ratio, reflecting the high-density behavior of
nuclear symmetry energy. This indicates that nucleons around low
rapidity $y/y_{beam}$= 0.2 are from high-density region of
compressed nuclear matter, and this part of emitted nucleons has a
large number (also as shown in Fig.\ \ref{pd}). Thus the free
\emph{n/p} ratios at large or small rapidities may reflect low or
high density behavior of symmetry energy. And because the emitted
nucleons are mainly at low-angle regions (shown in Fig.\
\ref{cosd}), we deduce that the better way to probe the effects of
symmetry energy on free neutron to proton ratio \emph{n/p} is
probing the emitted nucleons at large rapidities and low-angle
regions in laboratory system in the central reaction
$^{40}$Ca+$^{124}$Sn at 140 MeV/nucleon.

Fig.\ \ref{rnp2} shows the neutron to proton ratio (\emph{n/p}) of
the emitted free nucleons as a function of \emph{kinetic energy}
with $x=1$ and 0 in laboratory system in the central reaction
$^{40}$Ca+$^{124}$Sn at 140 MeV/nucleon. We can see that the stiff
symmetry energy ($x=0$) corresponds large \emph{n/p} at higher
kinetic energies. At lower kinetic energies the soft symmetry
energy ($x=1$) corresponds larger \emph{n/p}. From Fig.\
\ref{rnp2} and Fig.\ \ref{rnp}, we can again deduce that nucleons
at small rapidities mainly come from the compressed nuclear
matter, which have large kinetic energies, their \emph{n/p} ratios
reflect high density behavior of nuclear symmetry energy. Nucleons
at large rapidities mainly come from the low-density nuclear
matter, which have no very large kinetic energies, their
\emph{n/p} ratios reflect low density behavior of nuclear symmetry
energy. From above discussions, in the central reaction
$^{40}$Ca+$^{124}$Sn at 140 MeV/nucleon, to probe the high density
behavior of nuclear symmetry energy, we can probe nucleons at very
high kinetic energies and at mid-rapidities. To probe the low
density behavior of nuclear symmetry energy, we can probe nucleons
at not very high kinetic energies and at large rapidities.
Therefore, to probe the symmetry energy, it is better to give the
kinetic distribution and the rapidity distribution of emitted
nucleons at the same time.

\section{Summary}

Based on the semiclassical Boltzmann-Uehling-Uhlenbeck (BUU)
transport model, we studied the effects of symmetry energy on the
free neutron to proton ratio in the central reaction
$^{40}$Ca+$^{124}$Sn at 140 MeV/nucleon in the laboratory system.
It is found that at large rapidities free nucleon's neutron to
proton ratio is quite sensitive to the symmetry energy. And we
also find that free nucleons are mainly emitted at low-angle
regions and at \emph{zero} longitudinal momentum ($p_{z}$) in the
laboratory system. The maximal longitudinal momentum at positive
momentum part reached is about 0.9 GeV/c, but the maximal
longitudinal momentum at negative momentum part reached is only
about 0.3 GeV/c. And more nucleons are emitted at positive
momentum part (mainly in the range of $0<p_{z}<0.3 GeV/c$) than at
the negative part. The free \emph{n/p} ratios at small or large
rapidities may reflect high or low density behavior of nuclear
symmetry energy. These information is useful for related
experiments relevant to symmetry energy studies.

\section*{Acknowledgments}

The work is supported in part by the National Science Foundation
under Grant No.(10947109, 10875151, 11175219, 10740420550), the
Knowledge Innovation Project (KJCX2-EW-N01) of Chinese Academy of
Sciences, the Major State Basic Research Developing Program of
China under No. 2007CB815004, the CAS/SAFEA International
Partnership Program for Creative Research Teams (CXTD-J2005-1).
the Fundamental Research Fund for Physics and Mathematic of
Lanzhou University LZULL200908, the Fundamental Research Funds for
the Central Universities lzujbky-2010-160.

\end{CJK*}

\end{document}